\documentclass[12pt]{article}
\title{Nonclassical symmetries as special solutions of heir-equations}
\author{M C Nucci}
\date{Dipartimento di Matematica e Informatica\\
 Universit\`a di Perugia, 06123 Perugia, Italy\\
 E-mail: nucci@unipg.it}

\begin{document}
\baselineskip =24pt
 \maketitle
\begin{abstract}
In Nucci (1994), we have found that iterations of the nonclassical
symmetries method give rise to new nonlinear equations, which
inherit the Lie point symmetry algebra of the given equation. In
the present paper, we show that special solutions of the
right-order heir-equation correspond to classical and nonclassical
symmetries of the original
 equations. An infinite number of nonlinear equations
 which possess  nonclassical symmetries are derived.
\end{abstract}


\newpage

\section{Introduction}

The most famous and established method for finding exact solutions
of differential equations is the classical symmetries method
(CSM), also called group analysis, which originated in 1881
 from the pioneering work of Sophus Lie \cite{Lie}.
 Many good books have been dedicated to this
subject and its generalizations \cite{Ame72}, \cite{BluC74},
\cite{Ovs}, \cite{Olv}, \cite{BluK}, \cite{RogA}, \cite{Ste},
\cite{Hil}.

The nonclassical symmetries method (NSM) was introduced in 1969 by
Bluman and Cole \cite{BluC69} in order to obtain new exact
solutions of
 the linear heat equation, i.e. solutions not deducible from the
 CSM.
 The NSM
 consists of adding the invariant surface condition  to the given equation, and then
 applying the CSM.
 The main difficulty of this approach is that
 the determining equations are no longer linear.
 On the other hand, the NSM may yield more solutions than the CSM.
The NSM has been successfully applied to various equations
\cite{LevW}, \cite{ClaW}, \cite{peter}, \cite{ClaM}
\cite{GruT}\footnote{Just to cite some
 of numerous papers on this subject.},
 for the purpose of finding new exact solutions.

 Galaktionov \cite{Galak} and King \cite{Kin}
 have found exact solutions of certain evolution equations
 which apparently do not seem to be derived by either the CSM or NSM. In
\cite{itera}, we have  shown how these solutions can be obtained
by iterating the NSM. A special case of the NSM  generates a new
nonlinear equation (the so-called $G$-equation \cite{sgabt}),
which inherits the prolonged symmetry algebra of the original
equation. Another special case of the NSM is then applied to this
heir-equation to generate another heir-equation, and so on.
 Invariant solutions of these heir-equations are exactly the solutions
derived in \cite{Galak}, and \cite{Kin}.

In this paper, we show that the difficulty of finding nonclassical
symmetries can be overcome by determining the right-order
heir-equation, and  looking for a particular solution which has an
a priori known form. Both classical and nonclassical symmetries
can be found in this way. Therefore, our method may give an answer
to the question ``How can one establish a priori if a given
equation admits nonclassical symmetries?". We limit our analysis
to single evolution equations in two independent variables.  In
the present paper, we will not deal with systems, although
heir-equations for systems were introduced in \cite{allanu}.

In section 2, first we recall what heir-equations are, and then we
present our method. In section 3, some examples are given. In
section 4, we make some final comments.

 The use of a symbolic manipulator becomes imperative,
 because the heir-equations can be quite long:
one more independent variable is added at each iteration. We
employ  our own interactive REDUCE programs \cite{man} \cite{man2}
to generate the heir-equations.

\section{Heir-equations and outline of the method}

Let us consider an evolution  equation in two independent
variables and one dependent variable:
\begin{equation}
u_t=H(t,x,u,u_x,u_{xx},u_{xxx},\ldots) \label{pde}
\end{equation}
The invariant surface condition is given by:
\begin{equation}
V_1(t,x,u)u_t+V_2(t,x,u)u_x=F(t,x,u) \label{invsurf}
\end{equation}
Let us take the case with $V_1=0$ and $V_2=1$, so that
(\ref{invsurf}) becomes\footnote{We have replaced $F(t,x,u)$ with
$G(t,x,u)$ in order to avoid  any ambiguity in the following
discussion.}:
\begin{equation}
u_x=G(t,x,u)  \label{parG}
\end{equation}
 Then, an
equation for $G$ is easily obtained. We call this equation
$G$-equation \cite{sgabt}. Its invariant surface condition is
given by:
\begin{equation}
\xi_1(t,x,u,G)G_t+\xi_2(t,x,u,G)G_x+\xi_3(t,x,u,G)G_u=\eta(t,x,u,G)
\label{invsG1} \end{equation} Let us consider the case $\xi_1=0$,
$\xi_2=1$, and $\xi_3=G$, so that (\ref{invsG1}) becomes:
\begin{equation}
G_x+GG_u=\eta(t,x,u,G)
\end{equation}
Then, an equation for $\eta$ is derived. We call this equation
$\eta$-equation. Clearly:
\begin{equation}
 G_x+GG_u\equiv u_{xx}\equiv \eta
\end{equation}
We could keep iterating to obtain the $\Omega$-equation, which
corresponds to:
\begin{equation}
\eta_x+G\eta_u+\eta\eta_G \equiv u_{xxx} \equiv
\Omega(t,x,u,G,\eta)
\end{equation}
the $\rho$-equation, which corresponds to:
\begin{equation}
\Omega_x+G\Omega_u+\eta\Omega_G+\Omega\Omega_{\eta} \equiv
u_{xxxx}
 \equiv \rho(t,x,u,G,\eta,\Omega)
\end{equation}
and so on. Each of these equations inherits the symmetry algebra
of the original equation, with the right prolongation: first
prolongation for the $G$-equation, second prolongation for the
$\eta$-equation, and so on. Therefore, these equations are named
heir-equations.

This iterating method yields both partial symmetries as given by
Vorobev in \cite{Vor1}, and differential constraints as given by
Olver \cite{Olv2}. Also, it should be noticed that the
$u_{\underbrace{xx\cdots}_n}$--equation
 of (\ref{pde}) is just one of
many possible  $n$-extended equations as defined by Guthrie in
\cite{Gut}.

More details can be found in \cite{itera}.

Now, we describe the method that allows one to find nonclassical
symmetries of (\ref{pde}) by using a suitable heir-equation. For
the sake of simplicity, let us assume that the highest order
$x$-derivative appearing in (\ref{pde}) is two, i.e.:
\begin{equation}
u_t=H(t,x,u,u_x,u_{xx}) \label{pdeo2}
\end{equation}
First, we use (\ref{pdeo2}) to replace $u_t$ into (\ref{invsurf}),
with the condition $V_1=1$, i.e.:
\begin{equation}
H(t,x,u,u_x,u_{xx})+V_2(t,x,u)u_x=F(t,x,u) \label{cismod0}
\end{equation}
  Then, we generate the
$\eta$-equation with $\eta=\eta(x,t,u,G)$, and replace $u_x=G$,
$u_{xx}=\eta$ in (\ref{cismod0}), i.e.:
\begin{equation}
H(t,x,u,G,\eta)=F(t,x,u)-V_2(t,x,u)G \label{cismod}
\end{equation}
For Dini's theorem, we can isolate $\eta$ in  (\ref{cismod}),
e.g.:
\begin{equation}
\eta=\left[h_1(t,x,u,G)+F(t,x,u)-V_2(t,x,u)G\right]h_2(t,x,u,G)
\label{etasol}
\end{equation}
where $h_i(t,x,u,G) (i=1,2)$ are known functions.
 Thus, we have obtained
a particular solution of $\eta$ which must yield  an identity if
replaced into the $\eta$-equation. The only unknowns are
$V_2=V_2(t,x,u)$ and $F=F(t,x,u)$. Let us recall to the reader
that there are two sorts of nonclassical symmetries, those where
in (\ref{invsurf})
 the infinitesimal $V_1$ is non-zero, and those where it is zero \cite{ClaM}. In the first
case, we can assume without loss of generality that $V_1=1$, while
in the second case we can assume $V_2=1$, which corresponds to
generate the $G$-equation. If there exists a nonclassical
symmetry\footnote{Of course, we mean one such that  $V_1\neq 0$,
i.e. $V_1=1$ }, our method will recover it. Otherwise, only the
classical symmetries will be found. If we are interested in
finding only nonclassical symmetries, then we should impose $F$
and $V_2$ to be functions only of the dependent variable $u$.
Moreover, any such solution should be singular, i.e. should not
form a group.

 If we are dealing with a
third order equation, then we need to construct the heir-equation
of order three, i.e. the $\Omega$-equation. Then, a similar
procedure will yield a particular solution of the
$\Omega$-equation given by a formula similar to:
\begin{equation}
\Omega=\left[h_1(t,x,u,G,\eta)+F(t,x,u)-V_2(t,x,u)G\right]h_2(t,x,u,G,\eta)
\label{Omegasol}
\end{equation}
where $h_i(t,x,u,G,\eta) (i=1,2)$ are known functions.

In the case of a fourth order equation, we need to construct the
heir-equation of order four, i.e. the $\rho$-equation. Then, a
similar procedure will yield a particular solution of the
$\rho$-equation given by a formula  similar to:
\begin{equation}
\rho=\left[h_1(t,x,u,G,\eta,\Omega)+F(t,x,u)-V_2(t,x,u)G\right]h_2(t,x,u,G,\eta,\Omega)
\label{rhosol}
\end{equation}
where $h_i(t,x,u,G,\eta,\Omega) (i=1,2)$ are known functions.

And so on.

\section{Some examples}
We present some examples of how the method works. We consider some
families of evolution equations of second and third order. For
each of them, we derive the corresponding heir-equations up to the
appropriate order. Then, we look for the particular solution which
yields nonclassical symmetries. We would like to underline how
easy this method is in comparison with the existing one. The only
difficulty consists is deriving the heir-equations, which become
longer and longer. However, they can be automatically determined
by using any computer algebra system.
\subsection{Example 1:}
\begin{center}
 \framebox[4cm]{$u_t=u_{xx}+R(u,u_x)$}
\end{center}
 Let us consider the following family of second order evolution
 equations:
 \begin{equation}
 u_t=u_{xx}+R(u,u_x)\label{ex1}
  \end{equation}
 with $R(u,u_x)$ a known function of $u$ and $u_x$.
Famous equations known to possess  nonclassical symmetries  belong
to (\ref{ex1}):
 Burgers' \cite{Ame72}, Fisher's \cite{ClaM93}, real Newell-Whitehead's \cite{NewW},
 Fitzhugh-Nagumo's \cite{peter},
  and Huxley's equation \cite{ClaM93}, \cite{phil}.\\
  The $G$-equation of (\ref{ex1}) is:
\begin{eqnarray}
R_G(G G_u +  G_x) +  G R_u + 2 G_{xu} G + G_{uu} G^2 - G_u R - G_t
+ G_{xx}=0
\end{eqnarray}
The $\eta$-equation of (\ref{ex1}) is:
\begin{eqnarray}
2  R_{uG}  \eta  G + R_{GG}  \eta^2 + R_G \eta_x + G R_G \eta_u  +
R_{uu}  G^2 -G R_u \eta_G  \nonumber \\+ R_u \eta + 2 \eta_{xG}
\eta + 2 \eta_{uG} \eta G + \eta_{GG} \eta^2 -\eta_t + 2
\eta_{xu}G \nonumber \\ + \eta_{xx} + \eta_{uu}  G^2 - R \eta_u =0
\label{ex1etaeq}
\end{eqnarray}
The particular solution (\ref{etasol}) that  we are looking for
is:
\begin{equation}
\eta=-R(u,G)+F(t,x,u)-V_2(t,x,u) G \label{ex1eta}
\end{equation}
which replaced into (\ref{ex1etaeq}) yields an overdetermined
system in the unknowns $F$ and $V_2$ if $R(u,G)$ has a given
expression. Otherwise, after solving a first order linear partial
differential equation in $R(u,G)$, we obtain  that equation
(\ref{ex1}) may possess a nonclassical symmetry (\ref{invsurf})
with $V_1=1, V_2=v(u), F=f(u)$ if $R(u,u_x)$ has the following
form
\begin{equation}
R(u,u_x)={u_x\over f^2}\left( \left(- {{\rm d}f\over{\rm d}u} f
u_x + {{\rm d}v \over {\rm d}u}\right) f u_x^2 + \Psi(\xi) u_x^2 +
2 f^2 v - 3 f u_x v^2 + u_x^2 v^3 \right) \label{ex1.ru}
\end{equation}
with $f, v$ arbitrary functions of $u$, and $\Psi$ arbitrary
function of
 \begin{equation}
 \xi={f(u)\over u_x}-v(u) \label{xi}
\end{equation} This
means that infinitely many cases can be found.\\ Here we present
just three examples of (\ref{ex1.ru})
which are new, as far as we know.\\
Equation (\ref{ex1}) with $R(u,u_x)$ given by
\begin{equation}
R(u,u_x)=(2u_x + u^4){u_x \over u}
\end{equation} admits a nonclassical symmetry with  $v=u^3/2$ and
$f=-u^7/12$. It is interesting to notice that the corresponding
reduction leads to the solution of the following ordinary
differential equation in $u$ and $x$:
$$ u_{xx} =-2u_x^2/u-3u^3u_x/2-u^7/12$$
which is linearizable. In fact, it admits a Lie symmetry algebra
of dimension eight \cite{Lie2}. \\A second example is given by
equation (\ref{ex1}) with
\begin{eqnarray}
R(u,u_x)&=&u_x (16 \log(( - a_1^2 u^3 - 3 a_1 a_2 u^2 - 2 a_1 a_6
u^2 - 2 a_1 u_x u + 4 a_3 a_7 u \nonumber \\&&+ 4  a_4 a_7 - 4 a_5
a_7 u_x)/(4 a_7 u_x)) a_7^2 u_x^2+ a_1^5 u^7 + 7 a_1^4 a_2 u^6 + 4
a_1^4 a_6 u^6 \nonumber \\&&+ 15 a_1^3 a_2^2 u^5+ 16 a_1^3 a_2 a_6
u^5 - 8 a_1^3 a_3 a_7 u^5 - 8 a_1^3 a_4 a_7 u^4 \nonumber \\&&+ 4
a_1^3 a_6^2 u^5 + 9 a_1^2 a_2^3 u^4 + 12 a_1^2 a_2^2 a_6 u^4 - 32
a_1^2 a_2 a_3 a_7 u^4 \nonumber
\\&&- 32 a_1^2 a_2 a_4 a_7 u^3 + 4 a_1^2 a_2 a_6^2 u^4 - 16 a_1^2 a_3 a_6 a_7
u^4 - 16 a_1^2 a_4 a_6 a_7 u^3 \nonumber \\&& - 24 a_1 a_2^2 a_3
a_7 u^3 - 24 a_1 a_2^2 a_4 a_7 u^2 - 16 a_1 a_2 a_3 a_6 a_7
u^3\nonumber \\&& - 16 a_1 a_2 a_4 a_6 a_7 u^2 + 16 a_1 a_3^2
a_7^2 u^3 + 32 a_1 a_3 a_4 a_7^2 u^2 + 16 a_1 a_4^2 a_7^2 u
\nonumber \\&&+ 16 a_2 a_3^2 a_7^2 u^2 + 32 a_2 a_3 a_4 a_7^2 u +
16 a_2 a_4^2 a_7^2)/ (a_1^4 u^6 + 6 a_1^3 a_2 u^5 \nonumber
\\&&+ 4 a_1^3 a_6 u^5 + 9 a_1^2 a_2^2 u^4 + 12 a_1^2 a_2 a_6 u^4 -
8 a_1^2 a_3 a_7 u^4 \nonumber \\&&- 8 a_1^2 a_4 a_7 u^3 + 4 a_1^2
a_6^2 u^4 - 24 a_1 a_2 a_3 a_7 u^3 - 24 a_1 a_2 a_4 a_7 u^2
\nonumber \\&&- 16 a_1 a_3 a_6 a_7 u^3 - 16 a_1 a_4 a_6 a_7 u^2 +
16 a_3^2 a_7^2 u^2\nonumber \\&& + 32 a_3 a_4 a_7^2 u + 16 a_4^2
a_7^2)
\end{eqnarray}
where $a_j (j=1,7)$ are arbitrary constants. It admits a
nonclassical symmetry with
$$v= (a_1 u + a_2 + 2 a_6)/2$$ and $$f=( - a_1^2 u^3 - 3 a_1 a_2 u^2 - 2
a_1 a_6 u^2 + 4 a_3 a_7 u + 4 a_4 a_7)/4$$  A third example is
given by equation (\ref{ex1}) with
\begin{eqnarray}
R(u,u_x)&=&{u_x \over \sin(u)^2}\left(\cos(u)^3 u_x^2 - 3
\cos(u)^2 \sin(u) u_x + 2 \cos(u) \sin(u)^2 \right .\nonumber
\\&&\left. - \cos(u) \sin(u) u_x - \sin(u)^2 u_x^2 + \Psi(\xi) u_x^2\right)
\end{eqnarray}
where $\Psi$ is an arbitrary function of $\xi=( - \cos(u) u_x +
\sin(u))/u_x$. It admits a nonclassical symmetry with
$v=\cos(u)$ and $f=\sin(u)$. \\
 The subclass of equation (\ref{ex1}) with $R=r(u)$ was
considered in \cite{ClaM}, where classical and nonclassical
symmetries were retrieved. Just to show that our method is a lot
simpler than the existing one , we give the details of the
calculations in the case
$$ r(u)=-u^3-b u^2-c u-d
$$ which admits a nonclassical symmetry (\cite{ClaM}, Ansatz
4.2.1). We replace (\ref{ex1eta}) into (\ref{ex1etaeq}), which
becomes a third degree polynomial in $G$. The corresponding
coefficients (let us call them  $mm3, mm2, mm1, mm0$,
respectively)  must all become equal to zero. From $mm3=0$, we
obtain:
$$V_2=A_1(t,x)u+A_2(t,x)$$
 while $mm2=0$ yields:
 $$F= A_3(t,x)u + A_4(t,x) + {\partial A_1\over
\partial x} u^2 - A_1^2 u^3/3  - A_1 A_2 u^2$$
with $A_k(t,x) (k=1,4)$ arbitrary functions.
 Now, none of the remaining arbitrary functions depends on $u$. Since
 $mm1$ is a third degree polynomial in $u$, then its coefficients
 (let us call them $mm1k3, mm1k2, mm1k1, mm1k0$, respectively)
 must all become equal to zero.
 Equaling $mm1k3$ to zero yields two cases: either $A_1=0$ or $A_1^2=9/2$.
 If we assume $A_1=3/\sqrt{2}$, then $mm1k2=0, mm1k1=0, mm1k0=0$ lead to
 $ A_2=b/\sqrt{2}, A_4=-3c/2, A_3=-3d/2$, respectively. These
 values yield
 $mm0=0$. Thus, the nonclassical symmetry found in \cite{ClaM} is recovered, i.e.
  \begin{equation}
 V_2={b+3u \over \sqrt{2}},\;\;\;\;\;\;\;F={3\over 2} (-u^3-b u^2-c
 u-d)
 \end{equation}
 A similar result holds if we assume $A_1=-3/\sqrt{2}$: $F$ is the same, and
 $V_2=-(b+3u)/\sqrt{2}$. The case $A_1=0$ leads to either
 $V_2=1, F=0$ (trivial classical symmetry)
  or $d= - b(2b^2 - 9 c)/27$ with:
   $$V_2=A_2,\;\;\;\;\;\;\;\;\; F=- {1\over 3}(b + 3u)
 {\partial A_2\over \partial x}$$
 where $A_2(t,x)$ must satisfy:
 \begin{eqnarray}
 {\partial A_2\over \partial t} - 3 {\partial^2 A_2\over \partial x^2}
  + 2A_2 {\partial A_2\over \partial x}=0&&\nonumber \\
  3{\partial^3 A_2\over \partial x^3} - 3 A_2{\partial^2 A_2\over \partial x^2}
  + (b^2 - 3c){\partial A_2\over \partial x}=0&& \label{more}
\end{eqnarray}
Solving (\ref{more}) results into two more cases which  can be
found in \cite{ClaM} Table 2, fourth and fifth row, respectively.

\subsection{Example 2:} \begin{center}
 \framebox[5cm]{$u_t=u^{-2}u_{xx}+R(u,u_x)$}
 \end{center}
Let us consider another family of second order evolution
equations:
\begin{equation}
u_t=u^{-2}u_{xx}+R(u,u_x)\label{ex2}
\end{equation}
 The $G$-equation of (\ref{ex2}) is:
\begin{eqnarray}
R_G (G G_u  + G_x)u^3  + u^3 G R_u + 2u G G_{xu} + u G^2 G_x - u^3
R G_u  \nonumber \\ - 2 G^2 G_u  - u^3 G_t  + u G_{xx} - 2 G G_x=0
\end{eqnarray}
The $\eta$-equation of (\ref{ex2}) is:
\begin{eqnarray}
2 R_{uG} \eta G u^4 + R_{GG} \eta^2 u^4 + R_{G} \eta_{x} u^4 +
R_{G} \eta_{u} G u^4 + R_{uu} G^2 u^4\nonumber \\ - R_{u} \eta_{G}
G u^4 + R_u \eta u^4 + 2 \eta_{xG} \eta u^2 + 2 \eta_{uG} \eta G
u^2 + \eta_{GG} \eta^2 u^2 - 2 \eta_{G} \eta G u \nonumber \\-
\eta_{t} u^4 + 2 \eta_{xu} G u^2 + \eta_{xx} u^2 - 4 \eta_{x} G u
+ \eta_{uu} G^2 u^2 \nonumber \\- \eta_{u} R u^4 - 4 \eta_{u} G^2
u - 2 \eta^2 u + 6 \eta G^2=0 \label{ex2etaeq}
\end{eqnarray}
The particular solution (\ref{etasol}) that we are looking for is:
\begin{equation}
 \eta=\left[-R(u,G)+F(t,x,u)-V_2(t,x,u)  G \right]u^2
\label{ex2eta}
\end{equation}
which replaced into (\ref{ex2etaeq}) and imposing $V_2=v(u),
F=f(u)$ yields a first order linear partial differential equation
in $R(u,u_x)$. Then, equation (\ref{ex2}) may possess a
nonclassical symmetry (\ref{invsurf}) with $V_1=1, V_2=v(u),
F=f(u)$, if $R(u,u_x)$ has the following form
\begin{equation}
R(u,u_x)= -{\it v}u_x+{\it f}-{{\rm d}f\over{\rm du}}{u_x^2\over
uf} +\left({{\rm d}v\over{\rm d}u}f+ \Psi(\xi)\right)\,{{u_x}
^{3}\over  f^{2} u^{2}} \label{ex2.ru}
\end{equation}
with $f, v$ arbitrary functions of $u$, and $\Psi$ an arbitrary
function of the same $\xi$ as given in (\ref{xi}). This means that
infinitely many cases can be found.\\ Here we present just two
examples of (\ref{ex2.ru}) which are new, as far as we know. In
both examples, $\Psi$ is an arbitrary function of $\xi$ as
shown.\\ Equation (\ref{ex2}) with $R(u,u_x)$ given by
\begin{equation}
R(u,u_x)=2{u_x^3\over u^6}+\Psi\left ({\frac
{u^{5}}{u_x}}-u^{2}\right ){u_x^3\over u^{12}}-5 {u_x^2\over
u^3}-(1+u^2) u_x+u^5
\end{equation}
admits a nonclassical symmetry with  $v=u^2+1$ and $f=u^5$.\\
Equation (\ref{ex2}) with $R(u,u_x)$ given by
\begin{equation}
R(u,u_x)=\Psi\left({\frac {1}{uu_x}}- u \right)u_x^3+{u_x^3\over
u} +{u_x^2\over u^3}-uu_x+{1\over u}
\end{equation}
admits a nonclassical symmetry with  $v=u$ and $f=1/u$.\\
 Now we would like to show how our
method works with an equation which does not admit nonclassical
symmetries. Let us consider equation (\ref{ex2}) with $
R=-2u^{-3}u_x^2+1$ (\cite{Olv2} p.519), i.e.:
\begin{equation}
u_t=(u^{-2}u_x)_x+1 \label{olvp519-2}
\end{equation}
 Its $\eta$-equation  admits a solution of the type (\ref{ex2eta}) only
if:
 $$V_2={c_1+x\over c_3-2t},\;\;\;\;\;\;\;\;\;F={-2u\over c_3-2t}$$
where   $c_j (j=1,3)$ are arbitrary constants.
 It corresponds
to the three-dimensional Lie point symmetry algebra admitted by
(\ref{olvp519-2}). Nonclassical symmetries do not exist. However,
it is known that equation (\ref{olvp519-2}) admits  higher order
symmetries \cite{MSS}, which may be retrieved searching for
particular solutions of its higher order heir-equations, as we
conjecture in the final comments.

\subsection{Example 3:}
\begin{center}
 \framebox[5cm]{$u_t=u_{xxx} +R(u,u_x,u_{xx})$}
 \end{center}
 Let us consider the following family of third order evolution
 equations:
 \begin{equation}
 u_t=u_{xxx}+R(u,u_x,u_{xx})\label{ex3}
  \end{equation}
 with $R(u,u_x,u_{xx})$ a known function of $u$, $u_x$ and $u_{xx}$.
 We derive the $\Omega$-equation\footnote{Here we do not present
  anyone of the heir-equations due to their long
expression.} of (\ref{ex3}), and look for the particular solution
(\ref{Omegasol}), i.e.:
\begin{equation}
\Omega=-R(u,G,\eta)+F(t,x,u)-V_2(t,x,u) G \label{ex3Omega}
\end{equation}
which replaced into the $\Omega$-equation, and assuming $V_2=v(u),
F=f(u)$ yields the following first order linear partial
differential equation in $R(u,u_x,u_{xx})\equiv R(u,G,\eta)$
\begin{eqnarray}
R(u,G,\eta){\frac {d}{du}}f(u)+ {\frac {\partial }{\partial \eta
}}R(u,G,\eta) {G}^{3}{\frac {d^{2}}{d{u}^{2}}}v(u)\nonumber \\- {
\frac {\partial }{\partial \eta}}R(u,G,\eta) {G}^{2}{\frac {d^{
2}}{d{u}^{2}}}f(u)-3\,G\eta\,{\frac {d^{2}}{d{u}^{2}}}f(u)- {
\frac {\partial }{\partial \eta}}R(u,G,\eta) \eta\,{\frac {d}{d
u}}f(u)\nonumber\\- {\frac {\partial }{\partial G}}R(u,G,\eta) G{
\frac {d}{du}}f(u)+6\,{G}^{2}\eta\,{\frac
{d^{2}}{d{u}^{2}}}v(u)-3\,  {\frac {d}{du}}v(u)
v(u){G}^{2}\nonumber\\+3\,G {\frac {d}{du }}v(u)
f(u)-4\,R(u,G,\eta) {\frac {d}{du}}v(u) G\nonumber
\\+  {\frac {\partial }{\partial G}}R(u,G,\eta) {G}^{2}{ \frac {d}{du}}v(u)+3\, {\frac {\partial }{\partial
\eta}}R(u,G, \eta) G\eta\,{\frac {d}{du}}v(u)\nonumber\\- {\frac
{\partial }{
\partial u}}R(u,G,\eta) f(u)+3\,{\eta}^{2}{\frac {d}{du}}v(u)-{
G}^{3}{\frac {d^{3}}{d{u}^{3}}}f(u)+{G}^{4}{\frac
{d^{3}}{d{u}^{3}}}v( u)=0 \label{ex3.eqru}
\end{eqnarray}
 Thus, equation (\ref{ex3})
may possess a nonclassical symmetry (\ref{invsurf}) with $V_1=1,
V_2=v(u), F=f(u)$ if $R(u,u_x,u_{xx})\equiv R(u,G,\eta)$
 satisfies (\ref{ex3.eqru}).
Note that the complete integral of (\ref{ex3.eqru}) involves an
arbitrary function $\Phi=\Phi(\xi_1,\xi_2)$ of $\xi_1\equiv\xi$ as
given in (\ref{xi}) and
 \begin{equation}
 \xi_2=
{f(u)\over u_x^{3}} \left (u_{xx}\,f(u)-u_x^{2}{\frac
{d}{du}}f(u)+u_x^{3}{ \frac {d}{du}}v(u)\right ) \label{xi2}
\end{equation} This
means that infinitely many cases can be found.\\ Here we present
two classes of solutions of (\ref{ex3.eqru})
which have never been described, as far as we know.\\
Equation (\ref{ex3}) with $R(u,u_x,u_{xx})$ given by
\begin{eqnarray}
-3\,{\frac {{u}^{6}}{u_x}}-15\,{u}^{4}-6\,{\frac {u_{xx}\,{u}^
{3}}{u_x}}-12\,u_x{u}^{2}-\left (3\,u_x+12\,u_{xx}\right
)u+12\,{u_x}^{2}\nonumber
\\-3\,{ \frac {{u_{xx}}^{2}}{u_x}}+{\frac
{-6\,{u_x}^{2}+3\,u_{xx}\,u_x}{u}}+3\,{\frac {
{u_x}^{2}}{{u}^{2}}}-4\,{\frac {{u_x}^{3}}{{u}^{3}}}+3\,{\frac
{{u_x}^{3}}{{ u}^{4}}}-{\frac {{u_x}^{4}}{{u}^{5}}}+{\frac
{{u_x}^{4}}{{u}^{6}}}\nonumber
\\-{{u_x}^{4}\over {u}^ {9}} \,\Phi\left({\frac {u\left
(u_x+{u}^{2}\right )}{u_x}},-{\frac {{u}^{3} \left
(u_{xx}\,{u}^{3}-{u_x}^{3}-3\,{u_x}^{2}{u}^{2}\right
)}{{u_x}^{3}}}\right)
\end{eqnarray} admits a nonclassical symmetry with  $v=1-u$ and
$f=u^3$.\\ Equation (\ref{ex3}) with $R(u,u_x,u_{xx})$ given by
\begin{eqnarray}
{1\over 4\,{u}^{3}u_x\left (-2\,\sqrt {u}u_x+u\right )}
(12\,{u}^{3}u_{xx}\,{u_x}^{2}+12\,{u}^{9/2}{u_x}^{2}+6\,{u}
^{3/2}{u_x}^{5}-3\,{u_x}^{4}{u}^{2}\nonumber
\\-48\,{u_x}^{3}{u}^{4}+24\,{u}^{7/2}u_x{u_{xx}
}^{2}-24\,{u}^{5/2}{u_x}^{3}u_{xx}\nonumber
\\+4(u-2\sqrt
{u}{u_x})u_x^{5}\Phi\left(-{\frac {\sqrt {u}u_x-u}{u_x}},u\left
(u_{xx}\,u-{u_x}^{2}+{\frac {{u_x}^{3}}{2\sqrt {u}}}\right
){u_x}^{ -3}\right )\nonumber
\\-36\,{u}^{3}{u_x
}^{5}+64\,{u}^{7/2}{u_x}^{4}+8\,{u}^{5/2}{u_x}^{6}-12\,{u}^{4}{u_{xx}}^{2}\Big)
\end{eqnarray} admits a nonclassical symmetry with  $v=\sqrt{u}$ and
$f=u$.\\
Finally, we would like to show how our method works with a third
order evolution equation which does not admit nonclassical
symmetries. Let us consider the modified Korteweg-de Vries
equation (mKdV):
\begin{equation}
  u_t=u_{xxx}-6u^2u_x \label{mkdv}
\end{equation}
In \cite{sgabt}, the $G$-equation of (\ref{mkdv}) was derived:
\begin{eqnarray}
3 GG_{xxu}+3GG_uG_{xu}+3G_xG_{xu}+G^3G_{uuu}+3G^2G_{xuu}+
\nonumber \\
+3G^2G_uG_{uu}+3GG_xG_{uu}-G_t+G_{xxx}-6u^2G_x-12uG^2=0
\label{ex3Geq}
\end{eqnarray}
The $\eta$-equation of (\ref{mkdv}) is:
\begin{eqnarray}
 \eta^3  \eta_{GGG} + 3  \eta^2  \eta_{xGG} + 3 \eta^2 \eta_{uG}
+ 3  \eta^2  \eta_{uGG}  G + 3  \eta^2   \eta_G \eta_{GG}
\nonumber \\+ 3 \eta \eta_x  \eta_{GG} + 3  \eta  \eta_{xxG} + 3
\eta \eta_{xu} + 6 \eta \eta_{xuG}  G + 3  \eta   \eta_{xG} \eta_G
\nonumber \\ + 3 \eta \eta_u \eta_{GG} G + 3 \eta  \eta_{uu}  G +
3 \eta \eta_{uuG} G^2 + 3 \eta \eta_{uG} \eta_G G - 36 \eta u G -
\eta_t \nonumber \\+ 3 \eta_x \eta_{xG} + 3 \eta_x \eta_{uG} G - 6
\eta_x u^2 + \eta_{xxx} + 3  \eta_{xxu}  G + 3  \eta_{xuu}  G^2
\nonumber \\+ 3 \eta_{xG}  \eta_u  G
 +  3  \eta_u  \eta_{uG}
G^2 + \eta_{uuu}  G^3 + 12  \eta_G  u  G^2 - 12  G^3=0
\end{eqnarray}
The $\Omega$-equations of (\ref{mkdv}) is:
\begin{eqnarray}
6   \Omega_{xu\eta}  \Omega  G + 6   \Omega_{xG\eta} \Omega \eta +
3  \Omega_{xx\eta} \Omega + 3   \Omega_{x\eta} \Omega_{\eta}
\Omega + 3  \Omega_{x\eta}  \Omega_{x} + 3 \Omega_{x\eta}
 \Omega_{u} G \nonumber \\+ 3  \Omega_{x\eta}  \Omega_{G} \eta + 6
 \Omega_{uG\eta} \Omega \eta  G + 3  \Omega_{uu\eta}  \Omega
G^2 + 3   \Omega_{u\eta}  \Omega_{\eta} \Omega  G + 3
 \Omega_{u\eta}  \Omega_{x} G \nonumber \\+ 3  \Omega_{u\eta}
 \Omega_{u}  G^2 + 3  \Omega_{u\eta}  \Omega_{G} \eta  G + 3
 \Omega_{u\eta}  \Omega \eta + 3  \Omega_{GG\eta} \Omega
\eta^2 + 3  \Omega_{G\eta}  \Omega_{\eta} \Omega \eta \nonumber
\\+ 3 \Omega_{G\eta}  \Omega_{x} \eta + 3  \Omega_{G\eta}
  \Omega_{u} \eta G + 3  \Omega_{G\eta}  \Omega_{G} \eta^2 + 3
 \Omega_{G\eta} \Omega^2 +  \Omega_{\eta\eta\eta} \Omega^3\nonumber \\ + 3
 \Omega_{x\eta\eta} \Omega^2 + 3  \Omega_{u\eta\eta} \Omega^2  G +
3  \Omega_{G\eta\eta} \Omega^2 \eta + 3    \Omega_{\eta\eta}
 \Omega_{\eta} \Omega^2 + 3  \Omega_{\eta\eta}  \Omega_{x} \Omega\nonumber \\
+ 3  \Omega_{\eta\eta}  \Omega_{u}  \Omega G + 3 \Omega_{\eta\eta}
 \Omega_{G} \Omega \eta + 36  \Omega_{\eta} \eta u G + 12
 \Omega_{\eta} G^3 -  \Omega_{t} + 6  \Omega_{xuG} \eta G \nonumber \\+ 3
 \Omega_{xuu}  G^2+ 3  \Omega_{xu}  \eta + 3  \Omega_{xGG} \eta^2
  + 3  \Omega_{xG}  \Omega +  \Omega_{xxx} + 3
 \Omega_{xxu} G  \nonumber \\+ 3  \Omega_{xxG} \eta - 6  \Omega_{x} u^2
+ 3  \Omega_{uGG}  \eta^2 G + 3  \Omega_{uG}  \Omega G + 3
\Omega_{uG} \eta^2 +  \Omega_{uuu}  G^3   \nonumber \\+ 3
\Omega_{uuG} \eta G^2 + 3  \Omega_{uu} \eta G +  \Omega_{GGG}
\eta^3 + 3
 \Omega_{GG} \Omega  \eta + 12  \Omega_{G} u G^2  \nonumber \\- 48 \Omega u
G - 36 \eta^2 u - 72 \eta G^2=0 \label{ex3omegaeq}
\end{eqnarray}
The particular solution (\ref{Omegasol}) that  we are looking for
is:
\begin{equation}
\Omega=6u^2 G+F(t,x,u)-V_2(t,x,u)G \label{ex3sol}
\end{equation}
which replaced into (\ref{ex3omegaeq}) yields an overdetermined
system in the unknowns $F$ and $V_2$. It is very easy to prove
that nonclassical symmetries do not exist, a well-known result.
Instead, we obtain the classical symmetries admitted by
(\ref{mkdv}), i.e.:
\begin{equation}
V_2={ c_2  + x \over  c_1 + 3 t}\;\;\;\;\;\;\;\;\;\;\;\;\;
 F= - {u \over c_1 + 3 t}
\end{equation}
with $c_i (i=1,2)$ arbitrary constants.

\section{Final comments}
We have determined an algorithm which  is easier to implement than
the usual method to find nonclassical symmetries admitted by an
evolution equation in two independent variables. Moreover, one can
retrieve both classical and nonclassical symmetries with the same
method. Last but not least, we have shown that our method is able
to retrieve an infinite number of
equations admitting nonclassical symmetries.\\
 Using the heir-equations  raises many intriguing questions which we
 hope to address in future work:
 \begin{itemize}
\item Could an a priori knowledge of the existence of nonclassical
symmetries apart from classical be achieved by looking at the
properties of the right-order  heir-equation? We have shown that
our method leads to both classical and nonclassical symmetries.
Nonclassical symmetries could exist if we impose $F$ and $V_2$ to
be functions only of the dependent variable $u$  in either
(\ref{etasol}), or (\ref{Omegasol}), or (\ref{rhosol}), or
$\dots$. Of course, any such solution of $F$ and $V_2$ does not
yield a nonclassical symmetry, unless it is  isolated, i.e. does
not form a group.

\item What is integrability? The existence of infinitely many
higher order symmetries is one of the criteria \cite{MSS},
\cite{OSW}. In \cite{zhdanov}, we have shown that invariant
solutions of the heir-equations yield Zhdanov's conditional
Lie-B\"acklund symmetries \cite{Zhdanovpap}. Higher order
symmetries may be interpreted as special solutions of
heir-equations (up to which order?  see \cite{SanW}, \cite{OSW}).
Another criterion for integrability consists of looking for
B\"acklund transformations \cite{AblS}, \cite{RogS}. In
\cite{sgabt}, we have found that a nonclassical symmetry of the
$G$ equation (\ref{ex3Geq}) for the mKdV equation (\ref{mkdv})
gives the known B\"acklund transformation between (\ref{mkdv}) and
the KdV equation \cite{miura}. Another integrability test is the
Painlev\'e property \cite{WTC} which when satisfied leads to Lax
pairs (hence, inverse scattering transform) \cite{AblS},
B\"acklund transformations, and Hirota bilinear formalism
\cite{TabG}. In \cite{EstG}, the singularity manifold of the mkdV
equation (\ref{mkdv})  was found to be connected to an equation
which is exactly the $G$-equation (\ref{ex3Geq}). Could
heir-equations be the common link among all the integrability
methods?

\item In order to reduce a partial differential equation to ordinary differential equations,
one of the first things to do is find the admitted Lie point
symmetry algebra. In most instances, it is very small, and
therefore not many reductions can be obtained. However, if
heir-equations are considered, then many more ordinary
differential equations can be derived  using  the same Lie algebra
\cite{itera}, \cite{allanu}, \cite{MCN}. Of course, the
classification of all dimension subalgebras \cite{Win} becomes
imperative \cite{MCN}. In the case of known integrable equations
such as (\ref{mkdv}), it would be interesting to investigate which
ordinary differential equations result from using the admitted Lie
point symmetry algebra and the corresponding heir-equations. Do
all these ordinary differential equations possess the Painlev\'e
property (see the Painlev\'e conjecture as stated in \cite{ARS})?

\item In recent years, researchers often find  solutions of partial differential
equations which apparently do not come from any symmetry
reduction. Are the heir-equations the ultimate method which keeps
Lie symmetries at center stage?
\end{itemize}

\end{document}